\documentclass[aps,prl,twocolumn,tightenlines,superscriptaddress,showpacs,byrevtex]{revtex4}
\usepackage{graphicx}
\usepackage{dcolumn}
\usepackage{bm}
\usepackage{color}
\graphicspath{{eps/}}

\def\Dz {D^{0}}
\def\Db {\overline{D}{}^{\,0}}
\def\pip  {\pi^{+}}
\def\pim {\pi^{-}}

\def\mms {m_-^2}
\def\mps {m_+^2}
\def\dza {|{\Dz}\rangle}
\def\dba {|{\Db}\rangle}
\def\ks {K_S^0}

\def\decay{\Dz \to \ks\,\pip\pim}
\def\mass {m^{}_{\ks\pi\pi}}
\def\simlt{\mathrel{\lower2.5pt\vbox{\lineskip=0pt\baselineskip=0pt
          \hbox{$<$}\hbox{$\sim$}}}}
\def\ra{\!\rightarrow\!}

\begin{document}


\title{\boldmath {Measurement of $\Dz$-$\Db$ mixing in $\decay$ decays}}
\affiliation{Budker Institute of Nuclear Physics, Novosibirsk}
\affiliation{Chiba University, Chiba} \affiliation{University of
Cincinnati, Cincinnati, Ohio 45221}
\affiliation{The Graduate University for Advanced Studies, Hayama}
\affiliation{Hanyang University, Seoul} \affiliation{University of
Hawaii, Honolulu, Hawaii 96822} \affiliation{High Energy Accelerator
Research Organization (KEK), Tsukuba} \affiliation{Hiroshima
Institute of Technology, Hiroshima} \affiliation{University of
Illinois at Urbana-Champaign, Urbana, Illinois 61801}
\affiliation{Institute of High Energy Physics, Chinese Academy of
Sciences, Beijing} \affiliation{Institute of High Energy Physics,
Vienna} \affiliation{Institute of High Energy Physics, Protvino}
\affiliation{Institute for Theoretical and Experimental Physics,
Moscow} \affiliation{J. Stefan Institute, Ljubljana}
\affiliation{Kanagawa University, Yokohama} \affiliation{Korea
University, Seoul}
\affiliation{Kyungpook National University, Taegu}
\affiliation{Swiss Federal Institute of Technology of Lausanne,
EPFL, Lausanne} \affiliation{University of Ljubljana, Ljubljana}
\affiliation{University of Maribor, Maribor} \affiliation{University
of Melbourne, School of Physics, Victoria 3010} \affiliation{Nagoya
University, Nagoya} \affiliation{Nara Women's University, Nara}
\affiliation{National Central University, Chung-li}
\affiliation{National United University, Miao Li}
\affiliation{Department of Physics, National Taiwan University,
Taipei} \affiliation{H. Niewodniczanski Institute of Nuclear
Physics, Krakow} \affiliation{Nippon Dental University, Niigata}
\affiliation{Niigata University, Niigata} \affiliation{University of
Nova Gorica, Nova Gorica} \affiliation{Osaka City University, Osaka}
\affiliation{Osaka University, Osaka} \affiliation{Panjab
University, Chandigarh} \affiliation{Peking University, Beijing}
\affiliation{RIKEN BNL Research Center, Upton, New York 11973}
\affiliation{Saga University, Saga} \affiliation{University of
Science and Technology of China, Hefei} \affiliation{Seoul National
University, Seoul}
\affiliation{Sungkyunkwan University, Suwon} \affiliation{University
of Sydney, Sydney, New South Wales} \affiliation{Tata Institute of
Fundamental Research, Mumbai} \affiliation{Toho University,
Funabashi} \affiliation{Tohoku Gakuin University, Tagajo}
\affiliation{Tohoku University, Sendai} \affiliation{Department of
Physics, University of Tokyo, Tokyo} \affiliation{Tokyo Institute of
Technology, Tokyo} \affiliation{Tokyo Metropolitan University,
Tokyo} \affiliation{Tokyo University of Agriculture and Technology,
Tokyo}
\affiliation{Virginia Polytechnic Institute and State University,
Blacksburg, Virginia 24061} \affiliation{Yonsei University, Seoul}
 \author{L.~M.~Zhang}\affiliation{University of Science and Technology of China, Hefei} 
  \author{Z.~P.~Zhang}\affiliation{University of Science and Technology of China, Hefei} 
  \author{I.~Adachi}\affiliation{High Energy Accelerator Research Organization (KEK), Tsukuba} 
  \author{H.~Aihara}\affiliation{Department of Physics, University of Tokyo, Tokyo} 
  \author{V.~Aulchenko}\affiliation{Budker Institute of Nuclear Physics, Novosibirsk} 
  \author{T.~Aushev}\affiliation{Swiss Federal Institute of Technology of Lausanne, EPFL, Lausanne}\affiliation{Institute for Theoretical and Experimental Physics, Moscow} 
  \author{A.~M.~Bakich}\affiliation{University of Sydney, Sydney, New South Wales} 
  \author{V.~Balagura}\affiliation{Institute for Theoretical and Experimental Physics, Moscow} 
  \author{E.~Barberio}\affiliation{University of Melbourne, School of Physics, Victoria 3010} 
  \author{A.~Bay}\affiliation{Swiss Federal Institute of Technology of Lausanne, EPFL, Lausanne} 
% ...
\collaboration{The Belle Collaboration}

\begin{abstract}
We report a measurement of $\Dz$-$\Db$ mixing in $\decay$ decays
using a time-dependent Dalitz plot analysis. We first assume $CP$
conservation and subsequently allow for $CP$ violation. The results
are based on 540 fb$^{-1}$ of data accumulated with the Belle
detector at the KEKB $e^+e^-$ collider. Assuming negligible $CP$
violation, we measure the mixing parameters
$x=(0.80\pm0.29^{+0.09\,+0.10}_{-0.07\,-0.14})\%$ and
$y=(0.33\pm0.24^{+0.08\,+0.06}_{-0.12\,-0.08})\%$,
where the errors are statistical, experimental systematic, and
systematic due to the Dalitz decay model, respectively. Allowing for
$CP$ violation, we obtain the $CPV$ parameters
$|q/p|=0.86^{+0.30\,+0.06}_{-0.29\,-0.03}\pm0.08$ and
$\arg(q/p)=(-14^{+16\,+5\,+2}_{-18\,-3\,-4})^\circ$.
\end{abstract}


\maketitle

Mixing in the $D^0$-$\Db$ system is predicted to be very small in
the Standard Model (SM)~\cite{th1} and, unlike in $K^0$, $B^0$, and
$B^0_s$ systems, has eluded experimental observation. Recently,
evidence for this phenomenon has been found in $D^0\ra
K^+K^-/\pip\pim$ \cite{y_cp} and $D^0\ra K^+\pi^-$~\cite{kpi_BaBar}
decays. It is important to measure $D^0$-$\Db$ mixing in other decay
modes and to search for $CP$-violating ($CPV$) effects in order to
determine whether physics contributions outside the SM are present.
Here we study the self-conjugate decay $D^0\ra K^0_S\,\pi^+\pi^-$.

The time-dependent probability
of flavor eigenstates $\Dz$ and $\Db$ to mix to each other is
governed by the lifetime
  $\tau_{D^0}=1/\Gamma$, and the mixing parameters
$x=(m_1-m_2)/\Gamma$ and $y=(\Gamma_1-\Gamma_2)/2\Gamma$. The
parameters $m_1, m_2$ ($\Gamma_1,\Gamma_2$) are the masses (decay
widths) of the mass eigenstates $|D_{1,2}\rangle=p\dza\pm q\dba$,
and $\Gamma=(\Gamma_1+\Gamma_2)/2$.
The parameters $p$ and $q$ are complex coefficients satisfying
$|p|^2+|q|^2=1$.
Various $\Dz$ decay modes have been exploited to measure or
constrain $x$ and $y$  \cite{PDG_asner}. 
For $D^0\ra K^0_S\,\pi^+\pi^-$ decays, the time dependence of the
Dalitz plot distribution allows one to measure $x$ and $y$ directly.
This method was developed by CLEO~\cite{asner} using 9.0~fb$^{-1}$
of data; here we extend this method to a data sample 60 times
larger.

The decay amplitude at time $t$ of an initially produced $\dza$ or
$\dba$ can be expressed as
\begin{eqnarray}
{\cal M}(\mms,\mps,t)&=&{\cal A}(m_-^2,m_+^2)\frac{e_1(t)+e_2(t)}{2}
\nonumber\\&&+\frac{q}{p}\,\overline{\cal
A}(m_-^2,m_+^2)\frac{e_1(t)-e_2(t)}{2},  \nonumber \\
\overline{\cal M}(\mms,\mps,t)&=&\overline{\cal
A}(m_-^2,m_+^2)\frac{e_1(t)+e_2(t)}{2}\nonumber\\&&+\frac{p}{q}\,
{\cal A}(m_-^2,m_+^2)\frac{e_1(t)-e_2(t)}{2},
\label{eq1}
\end{eqnarray}
where ${\cal A}$ and $\overline{\cal A}$ are the amplitudes for
$\dza$ and $\dba$ decays as functions of the
invariant-masses-squared variables $m^2_\pm \equiv m^2(K^0_S\,\pi^\pm)$. 
The time dependence is contained in the terms
$e_{1,2}(t)=\exp[-i(m^{}_{1,2}-i\Gamma^{}_{1,2}/2)t]$.
Upon squaring ${\cal M}$ and $\overline{\cal M}$,
one obtains decay rates containing terms $\exp(-\Gamma t)\cos(x\Gamma t)$,
$\exp(-\Gamma t)\sin(x\Gamma t)$, and $\exp[-(1\pm y)\Gamma t]$.

We parameterize the $\ks\pip\pim$ Dalitz distribution following
Ref.~\cite{Anton}.
The overall amplitude as a function of $m^2_+$ and $m^2_-$
is expressed as a sum of quasi-two-body amplitudes (subscript $r$)
and a constant non-resonant term (subscript NR):
\begin{eqnarray}
{\cal A}(\mms,\mps)=\sum_r a_r e^{i\phi_r}{\cal
A}_r(\mms,\mps)+a^{}_{\rm NR}e^{i\phi_{\rm NR}},&&\\
 \overline{\cal
A}(\mms,\mps)=\sum_r \bar{a}_r e^{i\bar{\phi}_r}{\cal
A}_r(\mps,\mms)+\bar{a}^{}_{\rm NR}e^{i\bar{\phi}_{\rm NR}}.&&
\end{eqnarray}
The functions ${\cal A}^{}_r$ 
are
products of Blatt-Weisskopf form factors and relativistic
Breit-Wigner functions~\cite{kopp}.

The data were recorded by the Belle detector at the KEKB
asymmetric-energy $e^+e^-$ collider \cite{kek}. The Belle detector
\cite{belle} includes a silicon vertex detector (SVD), a central
drift chamber (CDC), an array of aerogel threshold Cherenkov
counters (ACC), a barrel-like arrangement of time-of-flight
scintillation counters (TOF), and an electromagnetic calorimeter.

We reconstruct $\Dz$ candidates via the decay chain $D^{*+}\to
\pi_s^+ \Dz$, $\Dz \to \ks \pi^+ \pi^-$~\cite{chargeconjugate}.
Here, $\pi_s$ denotes a low-momentum pion, the charge of which tags
the flavor of the neutral $D$ at production. The $\ks$ candidates
are reconstructed in the $\pi^+\pi^-$ final state; we require that
the pion candidates form a common vertex separated from the
interaction region and have an invariant mass within $\pm 10$
MeV/$c^2$ of $m^{}_{\ks}$. We reconstruct $\Dz$ candidates by
combining the $\ks$ candidate with two oppositely charged tracks
assigned as pions.
These tracks are required to have at least two SVD hits in both
$r$-$\phi$ and $z$ coordinates. A $D^{*+}$ candidate is
reconstructed by combining the $\Dz$ candidate with a low momentum
charged track (the $\pi_s^+$ candidate); the resulting $D^{*+}$
momentum in the $e^+e^-$
center-of-mass (CM) frame 
is required to be larger than 2.5 GeV/$c$ in order to eliminate
$B\overline{B}$ events and suppress combinatorial background.

The charged pion tracks are refitted to originate
  from a common vertex, which represents the decay point of the $\Dz$.
The $D^{*+}$ vertex
is taken to be the intersection of the $\Dz$ momentum vector with
the $e^+e^-$ interaction region.
The $\Dz$ proper decay time is calculated from the projection of the
vector joining the two vertices ($\vec{L}$)
onto the momentum vector: $t=\vec{L}\cdot(\vec{p}/p)
(m^{}_{D^0}/p)$.
The uncertainty in $t$ ($\sigma_t$) is calculated event-by-event,
and we require $\sigma_t < 1$~ps (for selected events, $\langle
\sigma_t\rangle\sim 0.2$~ps).

\begin{figure}[!hbtp]
\center
\includegraphics[bb=0 30 540 520, width=0.24\textwidth]{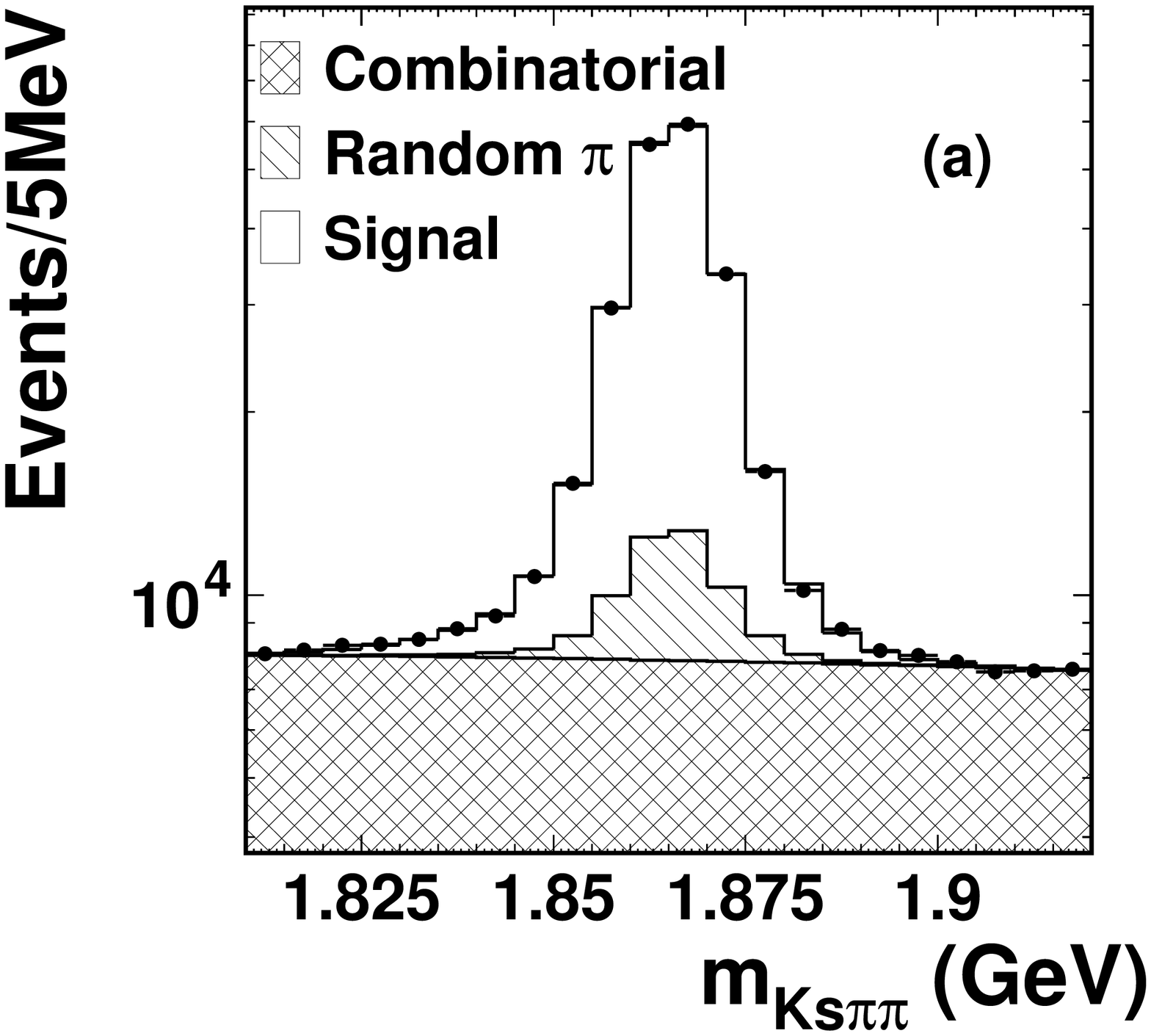}%
\includegraphics[bb=0 30 540 520,width=0.24\textwidth]{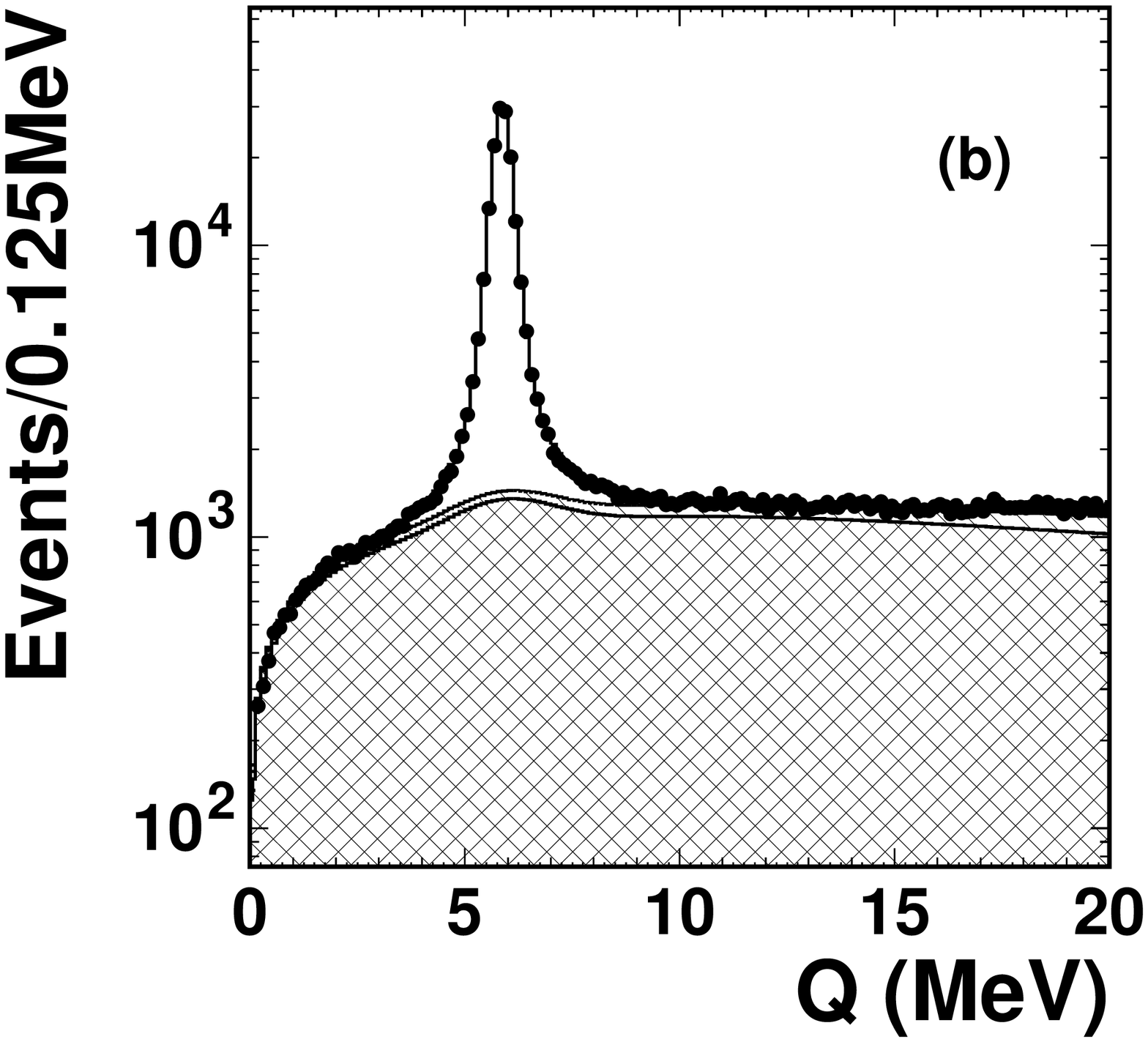}\break
\caption{\label{mqfig}The distribution of (a) $\mass$ with $0<Q<20$
MeV; (b) $Q$ with 1.81 GeV/$c^2<\mass< 1.92$ GeV/$c^2$. Superimposed
on the data (points with error bars) are projections of the
$\mass$-$Q$ fit.}
\end{figure}

The signal and background yields are determined from a
two-dimensional fit to the variables $\mass$ and $Q\equiv
(m_{\ks\,\pi\pi\pi^{}_s}\!-\mass\!-m^{}_{\pi})\cdot c^2$. The
variable $Q$ is the kinetic energy released in the decay and equals
only 5.9~MeV for $D^{*+}\ra \pi^{+}_s D^0$ decays. We parameterize
the signal shape by a triple-Gaussian function for $\mass$, and the
sum of a bifurcated Student $t$ distribution and a Gaussian function
for $Q$. The backgrounds are classified into two types: random
$\pi_s$ background, in which a random $\pi_s$ is combined with a
true $\Dz$ decay, and combinatorial background. The shape of the
$\mass$ distribution for the random $\pi_s$ background is fixed to
be the same as that used for the signal. Other background
distributions are obtained from Monte Carlo (MC) simulation. We
perform a two-dimensional fit to the measured $\mass$-$Q$
distributions in a wide range $1.81{\rm\
GeV}/c^2\!<\!\mass\!<\!1.92{\rm\ GeV}/c^2$ and $0\!<\!Q\!<\!20$~MeV.
We define a smaller signal region $|\mass-m_{\Dz}|<15$ MeV/$c^2$ and
$|Q-5.9$
MeV$|<1.0$ MeV, corresponding to $3\sigma$ intervals in 
these variables. In this region we find $534410\pm830$ signal events
and background fractions of 1\% and 4\% for the random $\pi_s$ and
combinatorial backgrounds, respectively.
The $\mass$ and $Q$ distributions are shown in
Fig.~\ref{mqfig} along with projections of the fit result.

For the events selected in the signal region we perform an unbinned
likelihood fit to the Dalitz plot variables $\mms$ and $\mps$, and
the decay time~$t$. For $\Dz$ decays, the likelihood function is
\begin{equation}
{\cal L}=\prod_{i=1}^{N_{\Dz}}\sum_{j}f_j(m_{\ks\pi\pi,i},Q_i) {\cal
P}_j(m_{-,i}^2,m_{+,i}^2,t_i)\,,
\end{equation}
where $j=\{\rm sig,rnd,cmb\}$ denotes the signal or background
components, and the index $i$ runs over $\Dz$ candidates. The event
weights $f_j$ are functions of $\mass$ and $Q$ and are obtained from
the $\mass$-$Q$ fit mentioned above.

The probability density function (PDF) ${\cal P}^{}_{\rm
sig}(m_{-}^2,m_{+}^2,t)$ equals $|{\cal M}(\mms,\mps,t)|^2$
convolved with the detector response. Resolution effects in
two-particle invariant masses are significant only for
$m^2_{\pi\pi}$.
The latter, and variation of the efficiency across the Dalitz plot,
are taken into account using the method described in
Ref.~\cite{Anton}. The resolution in decay time $t$ is accounted for
by convolving ${\cal P}^{}_{\rm sig}$ with a resolution function
consisting of a sum of three Gaussians with a common mean and widths
$\sigma_k=S_k\cdot \sigma_{t,i}$ ($k=1\!-\!3$). The scale factors
$S_k$ and the common mean are free parameters in the fit.

The random $\pi_s$ background contains real $\Dz$ and $\Db$ decays;
in this case the charge of the $\pi_s$ is uncorrelated with the
flavor of the neutral $D$.
Thus the ${\cal{P}}_{\rm rnd}$ PDF is taken to be $(1-f_{\rm
w})|{\cal M}(\mms,\mps,t)|^2+ f_{\rm w}|\overline{\cal
M}(\mms,\mps,t)|^2$, convolved with the same resolution function as
that used for the signal, where $f_{\rm w}$ is the wrong-tag
fraction. We measure $f_{\rm w}=0.452\pm0.005$ from fitting events
in the $Q$ sideband 3~MeV$<|Q-5.9{\rm\ MeV}|<14.1$~MeV.

For the combinatorial background, ${\cal{P}}_{\rm
    cmb}$ is the product of Dalitz-plot
and decay time PDFs. The latter
is parameterized as the sum of a delta function and an exponential
function convolved with a Gaussian resolution function. 
The timing and Dalitz PDF parameters are obtained from fitting
events in the mass sideband
30~MeV/$c^2$$<|\mass-m_{\Dz}|<55$~MeV/$c^2$.

\begin{figure}[!hbtp]
\center
\includegraphics[bb=0 30 540 520,width=0.24\textwidth]{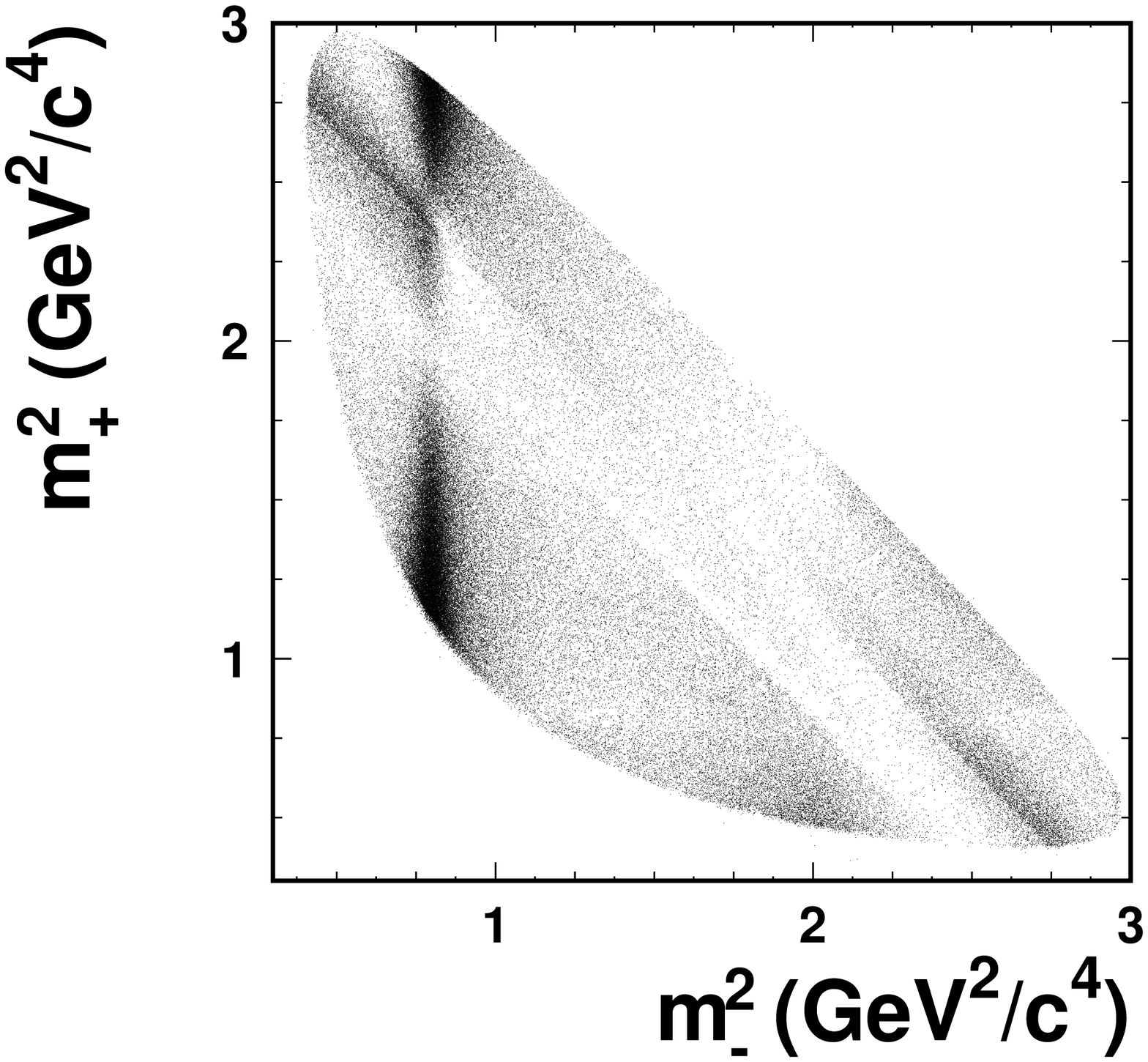}%
\includegraphics[bb=0 30 540 520,width=0.24\textwidth]{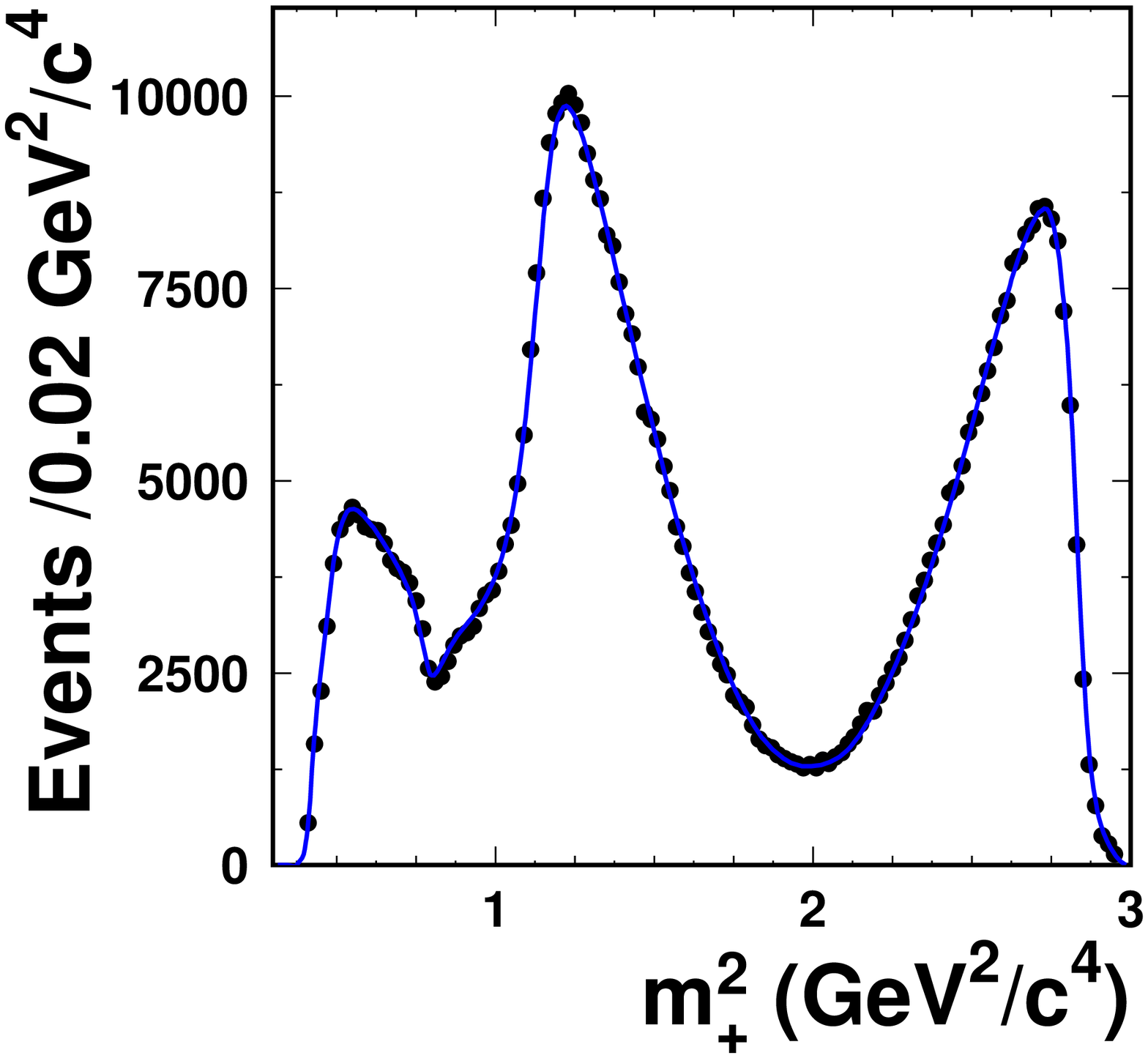}\break
\includegraphics[bb=0 30 540 520,width=0.24\textwidth]{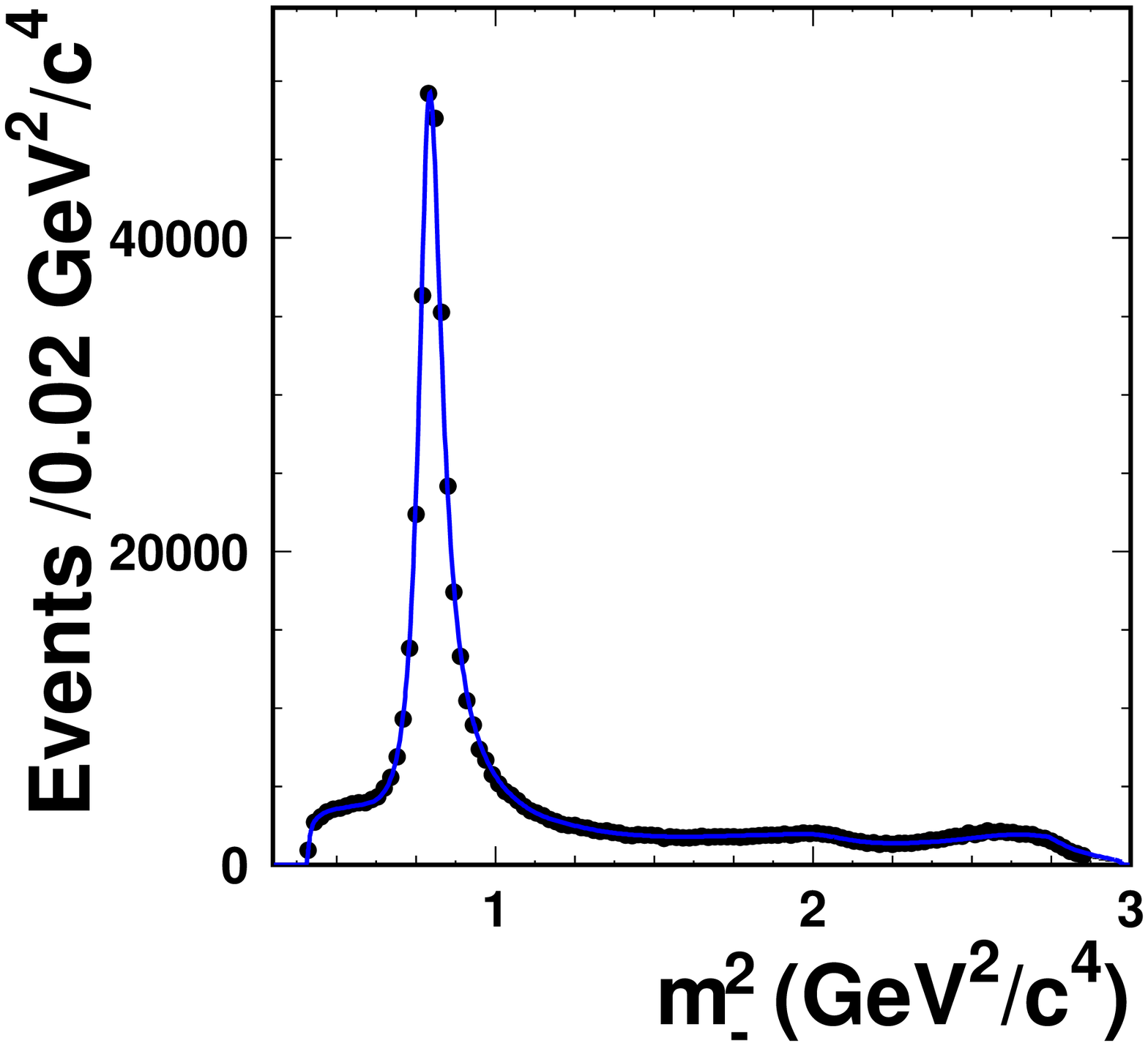}%
\includegraphics[bb=0 30 540 520,width=0.24\textwidth]{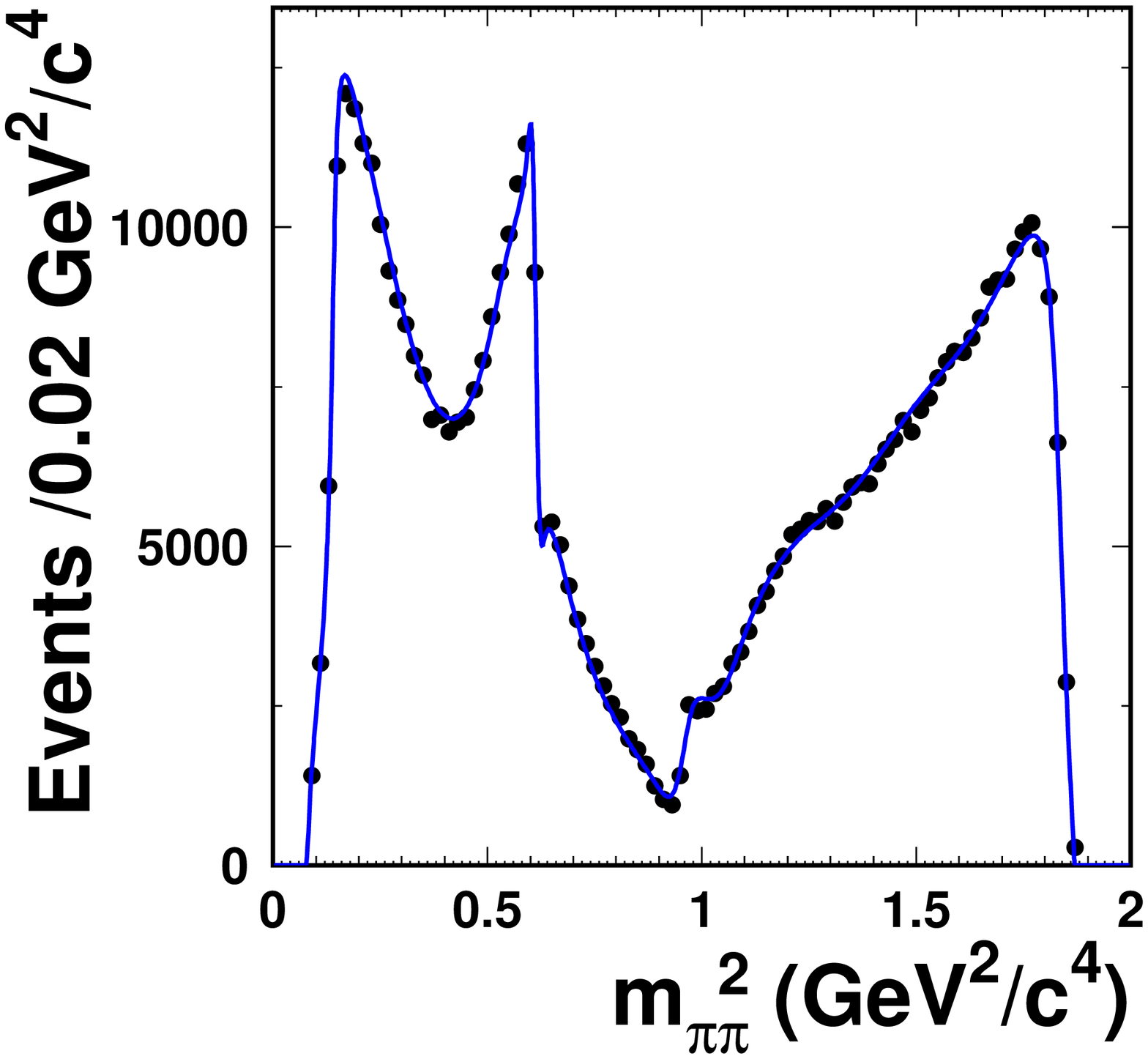}\break
\caption{\label{dfit_final} Dalitz plot distribution and the
projections for data (points with error bars) and the fit result
(curve). Here, $m^2_\pm$ corresponds to $m^{2}(\ks\pi^\pm)$ for
$\Dz$ decays and to $m^{2}(\ks\pi^\mp)$
 for $\Db$ decays.}
\end{figure}

\begin{figure}[!hbtp]
\center
\includegraphics[bb=0 0 540 520,width=0.35\textwidth]{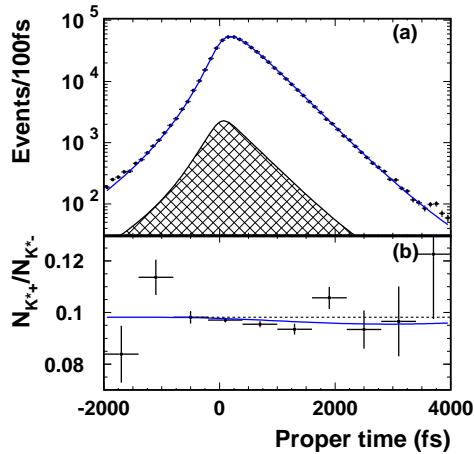}
\caption{\label{fig_t} (a) The decay-time distribution for events in
the Dalitz plot fit region for data (points with error bars), and
the fit projection for the $CP$-conservation fit (curve). The
hatched area represents the combinatorial background contribution.
(b) Ratio of decay-time distributions for events in the $K^*(892)^+$
and $K^*(892)^-$ regions.}
\end{figure}

The likelihood function for $\Db$ decays, $\overline{\cal L}$, has
the same form as ${\cal{L}}$, with ${\cal M}$ and $\overline{\cal
M}$ (appearing in ${\cal P}_{\rm sig}$ and ${\cal P}_{\rm rnd}$)
interchanged. To determine $x$ and $y$, we maximize the sum
$\ln{\cal L}+\ln\overline{\cal L}$.
Table \ref{table-final} lists the results from two separate fits. In
the first fit we assume $CP$ is conserved, i.e., $a=\bar{a}$,
$\phi=\bar{\phi}$, and
$p/q=1$. 
We fit all events in the signal region, where the free parameters
are $x$, $y$, $\tau^{}_{D^0}$, the timing resolution parameters of
the signal, and the Dalitz plot resonance parameters $a_{r(\rm NR)}$
and $\phi_{r(\rm NR)}$. The fit gives
$\tau^{}_{D^0}=(409.9\pm1.0)$~fs, which is consistent with the world
average~\cite{PDG}. The results for $a_r$ and $\phi_r$ for the 18
quasi-two-body resonances used (following the same model as in
Ref.~\cite{Anton}) and the NR contribution are listed in
Table~\ref{final-dlz}. The Dalitz plot and its projections, along
with projections of the fit result, are shown in
Fig.~\ref{dfit_final}. We estimate the goodness-of-fit of the Dalitz
plot through a two-dimensional $\chi^2$ test~\cite{Anton} and obtain
$\chi^2/ndf=2.1$ for $3653-40$ degrees of freedom ($ndf$). We find
that the main features of the Dalitz plot are well reproduced, with
some significant but numerically small discrepancies at peaks and
dips of the distribution in the very high $\mms$ region. The
decay-time distribution for all events, and the ratio of decay-time
distribution for events in the $K^*(892)^+$ and $K^*(892)^-$
regions, are shown in Fig.~\ref{fig_t}.

\begin{table}[!hbtp]
\renewcommand{\arraystretch}{1.2}
\center \caption {\label{table-final} Fit results and 95\% C.L.
intervals for $x$ and $y$, including systematic uncertainties. The
errors are statistical, experimental
systematic, and decay-model systematic, respectively. 
For the $CPV$-allowed case, there is another solution as described
in the text.}
\begin{tabular}{llcc}\hline
Fit case& Parameter & Fit result  & 95\% C.L. interval  \\\hline
No &$x(\%)$ & $0.80\,\pm 0.29\,^{+0.09\,+0.10}_{-0.07\,-0.14}$ & $(0.0,1.6)$ \\
$CPV$&$y(\%)$ & $0.33\,\pm 0.24\,^{+0.08\,+0.06}_{-0.12\,-0.08}$ &
$(-0.34,0.96)$
\\\hline
$CPV$&$x(\%)$ &$0.81\,\pm0.30\,^{+0.10\,+0.09}_{-0.07\,-0.16}$& $|x|<$1.6\\
&$y(\%)$ &$0.37\,\pm0.25\,^{+0.07\,+0.07}_{-0.13\,-0.08}$& $|y|<$1.04\\
&$|q/p|$ &$0.86^{+0.30\,+0.06}_{-0.29\,-0.03}\pm0.08$&-\\
&$\arg(q/p)(^\circ)$&$-14^{+16\,+5\,+2}_{-18\,-3\,-4}$&-\\\hline
\end{tabular}
\end{table}

\begin{table}[!hbtp]
\center \caption{\label{final-dlz} Fit results for Dalitz plot
parameters. The errors are statistical only. }
\begin{tabular}{lcrc}\hline\hline
Resonance & Amplitude& Phase ($^\circ$) &Fit fraction \\
\hline
$K^*(892)^-$&$1.629\pm0.006$&$134.3\pm0.3\phantom{.9}$&0.6227\\
$K_0^*(1430)^-$&$2.12\pm0.02$&$-0.9\pm0.8\phantom{.9}$&0.0724\\
$K_2^*(1430)^-$&$0.87\pm0.02$&$-47.3\pm1.2\phantom{.9}$&0.0133\\
$K^*(1410)^-$&$0.65\pm0.03$&$111\pm4\phantom{.9}$&0.0048\\
$K^*(1680)^-$&$0.60\pm0.25$&$147\pm29\phantom{.9}$&0.0002\\\hline
$K^*(892)^+$&$0.152\pm0.003$&$-37.5\pm1.3\phantom{.9}$&0.0054\\
$K_0^*(1430)^+$&$0.541\pm0.019$&$ 91.8\pm2.1\phantom{.9}$&0.0047\\
$K_2^*(1430)^+$&$0.276\pm0.013$&$-106\pm3\phantom{.9}$&0.0013\\
$K^*(1410)^+$&$0.33\pm0.02$&$-102\pm4\phantom{.9}$&0.0013\\
$K^*(1680)^+$&$0.73\pm0.16$&$103\pm11\phantom{.9}$&0.0004\\
\hline
$\rho(770)$& 1 (fixed)& 0 (fixed)$\phantom{.9}$&0.2111\\
$\omega(782)$&$0.0380\pm0.0007$&$115.1\pm1.1\phantom{.9}$&0.0063\\
$f_0(980)$&$0.380\pm0.004$&$-147.1\pm1.1\phantom{.9}$&0.0452\\
$f_0(1370)$&$1.46\pm0.05$&$98.6 \pm1.8\phantom{.9}$&0.0162\\
$f_2(1270)$&$1.43\pm0.02$&$-13.6\pm1.2\phantom{.9}$&0.0180\\
$\rho(1450)$&$0.72\pm0.04$&$41\pm7\phantom{.9}$&0.0024\\
$\sigma_1$&$1.39\pm0.02$&$-146.6\pm0.9\phantom{.9}$&0.0914\\
$\sigma_2$&$0.267\pm0.013$&$-157\pm3\phantom{.9}$&0.0088\\\hline
 NR&$2.36\pm0.07$&$155\pm2\phantom{.9}$&0.0615\\\hline\hline
\end{tabular}
\end{table}

For the second fit, we allow for $CPV$.  This introduces the
additional free parameters $|p/q|$, $\arg(p/q)$, $\bar{a}_{r(\rm
NR)}$ and $\bar{\phi}_{r(\rm NR)}$. The fit gives two solutions: if
\{$x$, $y$, $\arg(p/q)$\} is a solution, then \{$-x$, $-y$,
$\arg(p/q)+\pi$\} is an equally good solution. From the fit to data,
we find that the Dalitz plot parameters are consistent for the $\Dz$
and $\Db$ samples; hence we observe no evidence for direct $CPV$.
Results for $|p/q|$ and $\arg(p/q)$, parameterizing $CPV$ in mixing
and interference between mixed and unmixed amplitudes, respectively,
are also found to be consistent with $CP$ conservation. If we fit
the data assuming no direct $CPV$, the values for $x$ and $y$ are
essentially the same as those for the $CP$-conservation case, and
the values for the $CPV$ parameters are further constrained:
$|q/p|=0.95^{+0.22}_{-0.20}$ and $\arg(q/p)=(-2^{+10}_{-11})^\circ$.
A check with independent fits to the $\Dz$ and $\Db$ tagged samples
gives consistent results for $x$ ($y$): $0.58\%\pm0.41\%$
($0.45\%\pm 0.33\%$) and $1.04\%\pm0.41\%$ ($0.21\%\pm0.34\%$),
respectively.


We consider systematic uncertainties arising from both experimental
sources and from the $D^0\ra K^0_S\,\pi^+\pi^-$ decay model. We
estimate these uncertainties by varying relevant parameters by their
$\pm1\sigma$ errors and interpreting the change in $x$ and $y$ as
the systematic uncertainty due to that source. The main sources of
experimental uncertainty are the modeling of the background,
the efficiency, 
and the event
selection criteria. 
We vary the background normalization and timing parameters within
their uncertainties, and we also set $f_{\rm w}$ equal to its
expected value of 0.5 or alternatively let it float. To investigate
possible correlations between the Dalitz plot $(\mps,\mms)$
distribution and the $t$ distribution of combinatorial background,
the Dalitz plot distribution is obtained for three bins of decay
time; these PDFs are then used according to the reconstructed $t$ of
individual events.
We also try a uniform efficiency function, and we apply a
``best-candidate" selection to check the effect of the small
fraction of multiple-candidate events. We add all variations in $x$
and $y$ in quadrature to obtain the overall experimental systematic
error.

The systematic error due to our choice of
$\Dz\ra\ks\pip\pim$ decay model is evaluated as
follows.
We vary the masses and widths of the intermediate
resonances by their known uncertainties~\cite{PDG}, and
we also try fits with Blatt-Weisskopf form factors set
to unity and with no $q^2$ dependence in the Breit-Wigner
widths.
We perform a series of fits successively excluding intermediate
resonances that give small contributions ($\rho(1450),\
K^*(1680)^+$), and we also exclude the NR contribution. We account
for uncertainty in modeling of the $S$-wave $\pi\pi$ component by
using K-matrix formalism~\cite{kmatrix}. We include an uncertainty
due to the
effect of around 10-20\% bias in the amplitudes for the
$K^*(1410)^\pm$, $K_0^*(1430)^+$ and $K_2^*(1430)^+$ intermediate
states, which we observe in MC studies. Adding all variations in
quadrature gives the final results listed in
Table~\ref{table-final}.

\begin{figure}[!hbtp]
\center
\includegraphics[bb=0 30 540 520,width=0.4\textwidth]{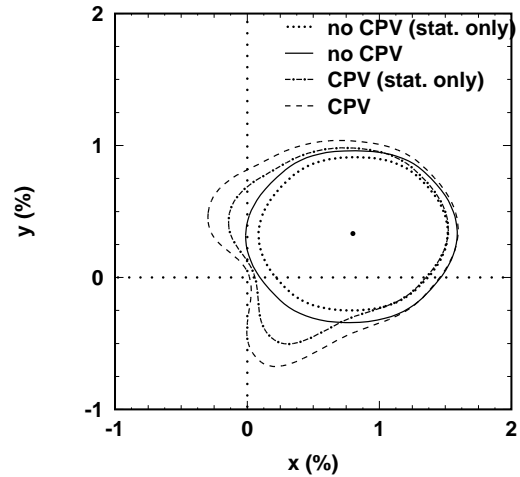}
\caption{\label{cn} 95\% C.L. contours for $(x,y)$: dotted (solid)
corresponds to statistical (statistical and systematic) contour for
no $CPV$, and dash-dotted (dashed) corresponds to statistical
(statistical and systematic) contour for the $CPV$-allowed case. The
point is the best-fit result for no $CPV$. }
\end{figure}

We obtain a 95\% C.L.\ contour in the
  $(x,y)$ plane by finding the
locus of points where $-2\ln{\cal L}$ increases by 5.99 units with
respect to the minimum value (i.e., $-2\Delta \ln{\cal L}$=5.99).
All fit variables other than $x$ and $y$ are allowed to vary to
obtain best-fit values at each point on the contour. To include
systematic uncertainty, we rescale each point on the contour by a
factor $\sqrt{1+r^2}$, where $r^2$ is a weighted average of the
ratios of systematic to statistical errors for $x$ and $y$, where
the weights depend on the position on the contour.
Both the statistical-only and overall contours for both the
$CPV$-allowed and the $CP$-conservation case are shown in
Fig.~\ref{cn}. We note that for the $CPV$-allowed case, the
reflection of these
contours through the origin $(0,0)$ are also allowed regions. 
Projecting the overall contour onto the $x,y$ axes gives the 95\%
C.L.\ intervals
listed in Table~\ref{table-final}.
After the systematics-rescaling procedure, the no-mixing point (0,0)
has a value $-2\Delta \ln{\cal L}=7.3$;
this corresponds to a C.L. of ~2.6\%. We have confirmed this value
by generating and fitting an ensemble of MC fast-simulated
experiments.

In summary, we have measured the $\Dz$-$\Db$ mixing parameters $x$
and $y$ using a Dalitz plot analysis of $\decay$ decays.
Assuming negligible $CP$ violation, we measure
$x=(0.80\pm0.29^{+0.09\,+0.10}_{-0.07\,-0.14})\%$ and
$y=(0.33\pm0.24^{+0.08\,+0.06}_{-0.12\,-0.08})\%$, where the errors
are statistical, experimental systematic, and decay-model
systematic, respectively. Our results disfavor the no-mixing point
$x\!=\!y\!=\!0$ with a significance of~$2.2\sigma$, while the one
dimensional significance for $x>0$ is $2.4\sigma$.
We have also searched for $CPV$
; we see no evidence for this and constrain the $CPV$ parameters
$|q/p|$ and $\arg(q/p)$.

\begin{acknowledgments}
We thank the KEKB group for excellent operation of the accelerator,
the KEK cryogenics group for efficient solenoid operations, and the
KEK computer group and the NII for valuable computing and
Super-SINET network support.  We acknowledge support from MEXT and
JSPS (Japan); ARC and DEST (Australia); NSFC and KIP of CAS (China);
DST (India); MOEHRD, KOSEF and KRF (Korea); KBN (Poland); MES and
RFAAE (Russia); ARRS (Slovenia); SNSF (Switzerland); NSC and MOE
(Taiwan); and DOE (USA).
\end{acknowledgments}


\begin{thebibliography}{999}

\bibitem{th1} A.~F.~Falk {\it et al.}, Phys.\ Rev.\ D {\bf 65}, 054034 (2002);
   I.~I.~Bigi, N.~Uraltsev, Nucl.\ Phys.\ B {\bf 592}, 92 (2001);
   A.~F.~Falk {\it et al.}, Phys.\ Rev.\ D {\bf 69},  114021 (2004);
   A.~A.~Petrov, Int.\ J.\ Mod.\ Phys.\ A{\bf 21}, 5686 (2006).

\bibitem{y_cp} M.~Stari\v{c} {\it et al.} (Belle
Collaboration), Phys. Rev. Lett. {\bf 98}, 211803 (2007).

\bibitem{kpi_BaBar} B. Aubert {\it et al.} (BaBar Collaboration),
  Phys. Rev. Lett. {\bf 98}, 211802 (2007).

\bibitem{PDG_asner} For a review see: D.~M.~Asner, $\Dz$- $\Db$ Mixing, in Ref. \cite{PDG}.




\bibitem{asner} D.~M.~Asner {\em et al.} (CLEO Collaboration), Phys.
Rev. D {\bf 72}, 012001 (2005) and arXiv: hep-ex/0503045v3.


\bibitem{Anton} A.~Poluektov {\em et al.} (Belle Collaboration),
  Phys. Rev. D {\bf 73}, 112009 (2006).

\bibitem{kopp} S.~Kopp {\em et al.} (CLEO Collaboration), Phys.
Rev. D {\bf 63}, 092001 (2001).


\bibitem{kek}S.~Kurokawa, E.~Kikutani \emph{et al.}, Nucl. Instrum. Methods Phys. Res. Sect. A
\textbf{499}, 1 (2003), and other papers in this volume.

\bibitem{belle}A.~Abashian \emph{et al.} (Belle Collaboration), Nucl. Instrum. Methods Phys.
Res. Sect. A \textbf{479}, 117 (2002); Z.~Natkaniec {\em et al.}
(Belle SVD2 Group), Nucl. Instrum. Methods Phys. Res. Sect. A
\textbf{560}, 1 (2006).

\bibitem{chargeconjugate} Charge conjugate decays are implied unless
  explicitly stated otherwise.

\bibitem{PDG} W.-M.~Yao {\em et al.} (Particle Data Group), J. Phys. G \textbf{33}, 1 (2006).

\bibitem{kmatrix} J.~M.~Link {\em et al.} (FOCUS Collaboration), Phys. Lett. B {\bf 585}, 200 (2004);
 B.~Aubert {\em et al.} (BaBar Collaboration), arXiv: hep-ex/0507101.

\end{thebibliography}
\end{document}